\newif\ifTWOCOLUMN \TWOCOLUMNfalse
\ifTWOCOLUMN
\documentclass[aps,prb,citeautoscript,twocolumn,floatfix]{revtex4}
\else
\documentclass[aps,prb,preprint]{revtex4}
\fi
\usepackage{bm}
\usepackage[dvipdfm]{graphicx}
\newcommand{\bi}{\bfseries\itshape}

\newcommand{\MH}{\gamma_{\rm{m}}}
\newcommand{\CC}{\Gamma_{\rm{c}}}
\newcommand{\CMC}{\Gamma_{\rm{c}} \times \gamma_{\rm{m}}}
\newcommand{\ask}{a_{\rm{sk}}}

\newcommand{\mnfegex}{{\rm Mn}_{1-x}{\rm Fe}_{x}{\rm Ge}}
\newcommand{\mnfesix}{{\rm Mn}_{1-x}{\rm Fe}_{x}{\rm Si}}
\newcommand{\fecosix}{{\rm Fe}_{1-x}{\rm Co}_{x}{\rm Si}}

\begin{document}
\title{Skyrmions with varying size and helicity in composition-spread helimagnetic alloys}
\author{K. Shibata$^{1*}$, X. Z. Yu$^2$, T. Hara$^3$, D. Morikawa$^2$, N. Kanazawa$^1$, K. Kimoto$^3$, S. Ishiwata$^1$, Y. Matsui$^3$ and Y. Tokura$^{1,2*}$}
\affiliation{
$^1$ Department of Applied Physics and Quantum-Phase Electronics Center (QPEC), University of Tokyo, Tokyo 113-8656, Japan,\\
$^2$ RIKEN Center for Emergent Matter Science (CEMS), Wako 351-0198, Japan,\\
$^3$ Surface Physics and Structure Unit, National Institute for Materials Science, Tsukuba 305-0044, Japan.
}
\maketitle
{\bf
The chirality, {\bi i.e.} left or right handedness, is an important notion in a broad range of science.
In condensed matter, this occurs not only in molecular or crystal forms but also in magnetic structures.
A magnetic skyrmion\cite{Bogdanov1989,Rossler2006,Muhlbauer2009,Yu2010,Yu2011,Tonomura2012,Seki2012}, a topologically-stable spin vortex structure, as observed in chiral-lattice helimagnets is one such example; the spin swirling direction (skyrmion helicity) should be closely related to the underlying lattice chirality via the relativistic spin-orbit coupling (SOC).
Here, we report on the correlation between skyrmion helicity and crystal chirality as observed by Lorentz transmission electron microscopy (TEM) and convergent-beam electron diffraction (CBED) on the composition-spread alloys of helimagnets $\mnfegex$ over a broad range ($x = 0.3$ -- $1.0$) of the composition.
The skyrmion lattice constant or the skyrmion size shows non-monotonous variation with the composition $x$, accompanying a divergent behavior around $x = 0.8$, where the correlation between magnetic helicity and crystal chirality is reversed.
The underlying mechanism is a continuous $x$-variation of the SOC strength accompanying sign reversal in the metallic alloys.
This may offer a promising way to tune the skyrmion size and helicity.
} 

The concept of skyrmion was originally introduced as a model to describe a state of nucleon\cite{Skyrme1962}, but is now extended to describe a spin configuration in quantum Hall\cite{Sondhi1993} and helimagnetic\cite{Bogdanov1989,Rossler2006,Muhlbauer2009,Yu2010,Yu2011,Tonomura2012,Seki2012} systems.
This particle-like nano-scale magnetic object, as schematized in Fig. 1a, is stabilized in the background of chiral crystal structure and constituent spin directions can wrap a sphere, {\it i.e.} solid angle of 4$\pi$ subtended by constituent spins, as characterized by a topological charge (skyrmion number) of $-1$.
Skyrmions have a typical size of 3--100 nm and they tend to crystallize mostly in a hexagonal lattice form, or sometime in a tetragonal or cubic lattice form\cite{Kanazawa2012}; we call such a magnetically ordered phase a skyrmion crystal (SkX).
SkX was at first identified by a small-angle neutron diffraction study on B20-type MnSi\cite{Muhlbauer2009} and then directly observed by Lorentz TEM in B20-type alloys Fe$_{0.5}$Co$_{0.5}$Si\cite{Yu2010}, FeGe\cite{Yu2011}, and MnSi\cite{Tonomura2012}.

Skyrmions and SkX have attracted much attention for emergent electromagnetic properties induced by their topological nature.
The skyrmion number corresponds directly to the gauge flux through quantum Berry phase, which manifests itself in topological Hall effects in SkX\cite{Lee2009,Neubauer2009,Kanazawa2011}.
SkX shows many other intriguing characteristics such as almost pinning-free motion with ultralow current density ($< 10^2$ $\mathrm{A/cm}^2$)\cite{Schulz2012,Yu2012}, electric polarization carried by skyrmions in an insulator\cite{Seki2012}, and circulating/breathing vortex-core motions as magnetic resonances\cite{Mochizuki2012,Onose2012}.
Despite of these interesting potentials of magneto-electronic functions, the crystal engineering toward the control of SkX structure itself, such as lattice constant, lattice form, and magnetic helicity, is not well established at all as compared with the wisdom for the control of conventional ferromagnetic domains.
Here, we report the composition control of the skyrmion size (SkX lattice constant) and the magnetic helicity in the mixed-crystal approach; we target the alloys of $\mnfegex$ with an intrinsic tendency of spinodal decomposition causing composition-spread domains, in which we map out the local SkX lattice constant, the local composition $x$ and the crystal chirality by combined means of Lorentz TEM, energy dispersive x-ray spectrometry (EDX), electron energy-loss spectroscopy (EELS), and CBED.
We have revealed not only a huge variation of the SkX lattice constant but also the helicity reversal of the skyrmion relative to the crystal chirality with variation of $x$, all caused by a continuous change of SOC strength accompanying its sign reversal in the mixed-crystal system.

\begin{figure}[t]
\begin{center}
\ifTWOCOLUMN
\includegraphics[width=7.5cm]{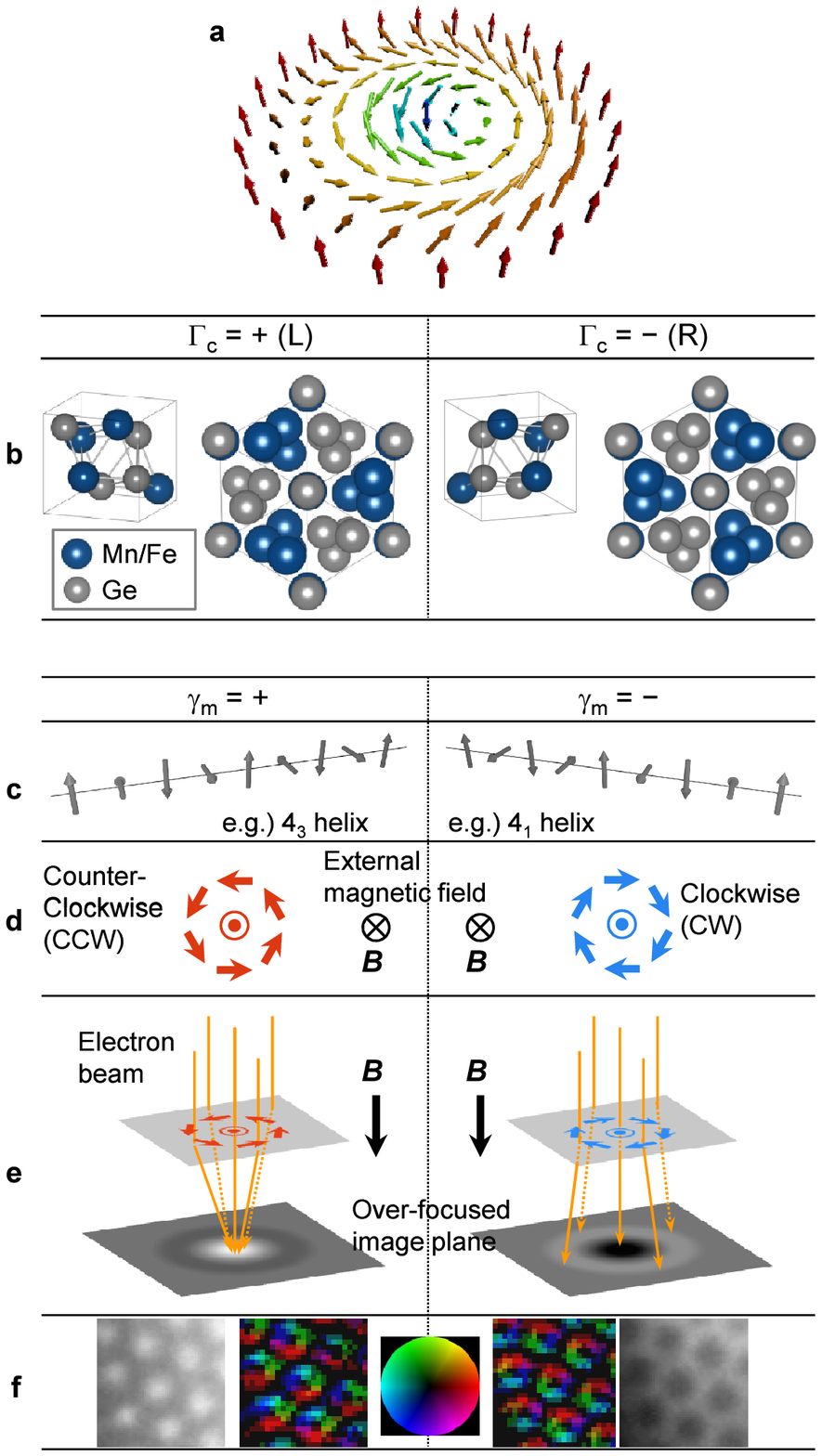}
\else
\includegraphics[width=8cm]{fig1.eps}
\fi
\caption{
{\bf Definitions of crystalline chirality $\CC$ and magnetic helicity $\MH$, and observation of skyrmions by Lorentz TEM.}
{\bf a}, A schematic illustration of the magnetic-moment configuration in a skyrmion.
{\bf b}, Unit cells and atom arrangements of left- and right-handed B20-type crystals viewed along [111] directions. 
{\bf c}, Magnetic-moment configurations of helical orders. 
{\bf d}, In-plane magnetic moment configurations of skyrmions, when magnetic field is applied downward. 
{\bf e}, Schematic views of how Lorentz TEM can be used to observe skyrmions and determine their helicity. 
{\bf f}, Over-focused Lorentz images of skyrmions in FeGe and corresponding lateral-magnetization-distribution maps obtained by the transport-of-intensity equation (TIE) analysis, respectively. 
The color images represent the direction and intensity of the in-plane magnetic moments (a color wheel shown at the center). 
}
\end{center}
\end{figure}
We show in Fig. 1 the definitions of crystalline and magnetic chiralities adopted in this paper.
The B20 structure belongs to the non-centrosymmetric space group of $\mathrm{P}2_13$ and can have enantiomers with right- and left-handed chiralities, as schematized by the projection along the [111] axis in Fig. 1b.
Here, we adopt the definition of crystallographic handedness by the chirality of carbon family atom (Ge) configuration.

In such a chiral-lattice helimagnet, the spin system is modeled by the effective Hamiltonian\cite{Landau2008},
\begin{equation}
H=\int d\bm{r}\left[
\frac{J}{2}(\nabla \bm{M})^2 + \alpha \bm{M} \cdot (\nabla \times \bm{M})
\right].
	\label{eq:hamiltonian}
\end{equation}

Here $\bm{M}$ is the spatially-varying magnetization, $J$ the ferromagnetic exchange interaction, $\alpha$ the Dzyaloshinskii-Moriya (DM) interaction constant, and $\bm{r}$ the three-dimensional position vector.
As the ground state, this Hamiltonian stabilizes a proper screw type helical magnetic structure in which the magnetic moment plane is perpendicular to the wave vector $\bm{q}$.
In this model, $\bm{q}$ has the magnitude proportional to $\alpha/J$.
Due to the asymmetric DM interaction, chirality of the spin helical structure, which we will hereafter call the magnetic helicity, depends on the sign of $\alpha$.
The sign of $\alpha$ is known to be determined by both the crystalline chirality ($\CC$) and the sign of SOC.
Figure 1c shows magnetic-moment configurations of the helical magnetic order.
When the propagation vector $\bm{q}$ is parallel (antiparallel) to $\bm{M}_1\times \bm{M}_2$ ($\bm{M}_1$ and $\bm{M}_2$ being the magnetic moments in order along the $\bm{q}$ direction), we call it the right- (left-) handed helix and define the magnetic helicity as $\MH=+ (-)$.
By this definition, the sign of $\MH$ corresponds one-to-one to that of $\alpha$.

Skyrmions in helimagnets emerge with applying an external magnetic field $\bm{B}$ normal to the thin-film plate.
They show a vortex-like configuration made up of the core magnetic moment anti-parallel to $\bm{B}$, the peripheral magnetic moment parallel to $\bm{B}$, and the transient-region magnetization swirling up in a clockwise (CW) or counter-clockwise (CCW) manner, as schematized in Fig. 1a.
Either of these two configurations is energetically favored by the sign of $\alpha$, and hence the corresponding $\MH$ is realized in a crystal domain with a fixed lattice chirality.
Here, we take the direction of $\bm{B}$ as vertically downward so as to match the present Lorentz TEM experimental configuration for the skyrmion observation.
Then, the in-plane magnetic-moment configuration of skyrmions is determined to be CCW (CW) for $\MH = + (-)$ as shown in Fig. 1d.
Lorentz TEM is a powerful method to visualize real-space magnetic-moment distribution of topological spin textures such as spin helices and skyrmions.
The incident electron beam is deflected by Lorentz force from the in-plane magnetic moments in the sample, and the spatial variation of the in-plane magnetization results in convergence (bright contrast) or divergence (dark contrast) on the defocused (under- or over-focused) image planes.
By using this method, a helical magnetic structure can be visualized as stripes\cite{Uchida2006}.
However, the Lorentz images for the proper screw structure cannot provide information on $\MH$ since only the in-plane component can be obtained with Lorentz TEM.
By contrast, Lorentz TEM images of skyrmions contain sufficient information for their helicities.
As demonstrated by previous Lorentz TEM studies\cite{Yu2010,Yu2011,Tonomura2012,Seki2012,Yu2012}, a  skyrmion can be visualized as a bright or dark spot.
As shown in Fig. 1e, the in-plane CCW (CW) magnetic-moment configuration deflects the electron beam and acts as a convex (concave) lens, resulting in a bright (dark) spot in the over-focused image plane.
Thus, in the over-focused Lorentz TEM image, the position of skyrmion and its chirality are visualized simultaneously as the spotty image and its contrast, respectively.
The observed over-focus Lorentz TEM images of FeGe are shown in Fig. 1f.
Reconstruction of the in-plane magnetic moment can be done by solving the transport-of-intensity equation (TIE), for the both over-focused and under-focused images\cite{Bajt2000,Ishizuka2005},
as also shown in Fig. 1f with use of a color wheel.

\begin{figure}[t]
\begin{center}
\ifTWOCOLUMN
\includegraphics[width=8cm]{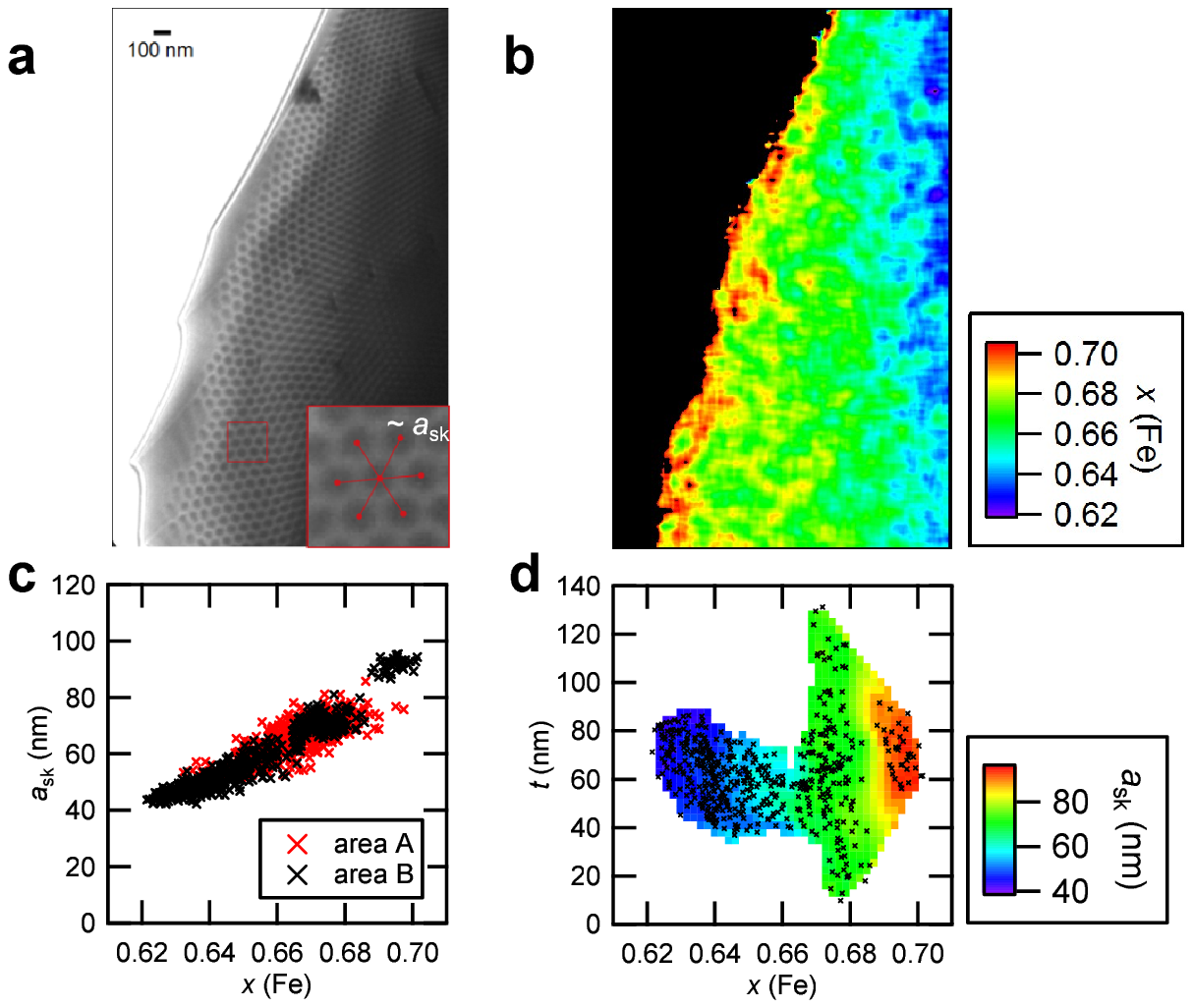}
\else
\includegraphics[width=12cm]{fig2.eps}
\fi
\caption{
{\bf Dependence of skyrmion size on composition $x$ obtained in a composition-spread micro-crystal of $\mnfegex$ ($x\sim0.7$).}
{\bf a}, Over-focused Lorentz images of SkX (area A). The SkX lattice constant $\ask$ at each skyrmion site is calculated by averaging distances from the 6 nearest neighbor sites (see the inset).
{\bf b}, The composition ($x$) distribution map obtained by EDX (area A).
{\bf c}, The composition $x$ dependence of $\ask$ (area A and B).
{\bf d}, The composition $x$ and thickness $t$ dependence of $\ask$ (area B).
}
\end{center}
\end{figure}
We show in Fig. 2 the dependence of skyrmion lattice constant (inter-skyrmion distance) $\ask$ on the composition $x$ and thickness $t$ for thinned-plate samples of $\mnfegex$.
Figure 2a presents an over-focused Lorentz TEM image of SkX in the sample with the nominal composition of $x = 0.7$.
The dark spots correspond to skyrmions.
Considerable variation of $\ask$ was observed within a microcrystalline domain.
A map of composition $x$ within the plate sample (area A, the same area shown in Fig. 2a) was obtained by EDX as shown in Fig. 2b.
An appreciate gradient of the composition was observed even in the micrometer-scale crystalline domain.
This is likely attributed to the spinodal decomposition in the mixed crystal between MnGe and FeGe; the actual quenching procedure during the sample preparation (see Methods) appears to produce a gradient of the composition $x$ (Fe) as steep as $\Delta x = 0.1$ over 1 $\mu$m as observed in Fig. 2b.
We have examined the correlation of $\ask$ with $x$ as follows: First, as SkX forms a triangular lattice, $\ask$ at each skyrmion site was calculated as the average of distances from the six nearest neighbors.
Then, the Lorentz-image and the EDX-composition map were superposed by fitting the edge shape of the sample and obtained the local composition on each skyrmion site.
The obtained composition dependence of $\ask$ is shown in Fig. 2c 
for area A shown in Fig. 2a or 2b and another region (area B, not shown).
For the both areas, a clear positive correlation is observed between $x$ and $\ask$.

In nanoscale magnets, a sample thickness may affect the stability of specific spin textures via magnetic anisotropy effect.
To check this, we have also investigated the thickness ($t$) dependence of $\ask$.
Mapping of the local thickness of the sample was taken by EELS in the area B.
The obtained map was again superposed with the Lorentz TEM image, and then the value of $t$ was extracted at each skyrmion site together with the local composition $x$ from the EDX mapping.
Figure 2d shows the contour map of $\ask$ in the plane of $t$ and $x$.
This shows the clear $x$-dependence of $\ask$, while its $t$-dependence is negligibly small in the present range of $t$ (= 20 -- 120 nm)  of the TEM sample.
Hereafter, therefore, we focus only on the $x$-dependence of the skyrmions.

\begin{figure*}[t]
\begin{center}
\includegraphics[width=10cm]{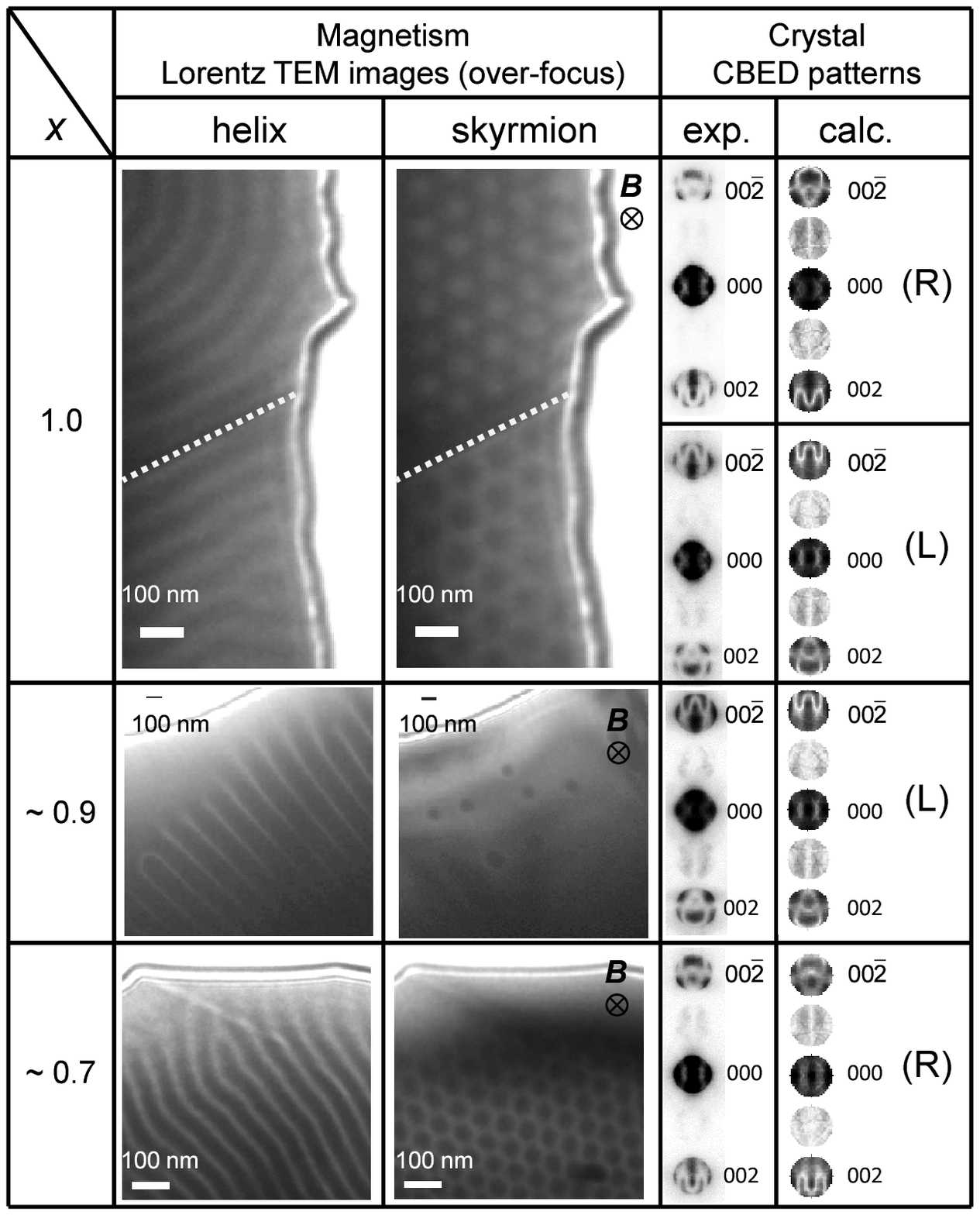}
\caption{
{\bf Lorentz TEM images of magnetic helix and skyrmion and CBED disk patterns used for determination of $\CC$.}
The over-focused Lorentz TEM images for helix and skyrmion were taken at zero magnetic field and at 100, 500, 200 mT and at 250, 95, 100 K, respectively, for the nominally $x$ = 1.0, 0.9 and 0.7 samples.
All the CBED patterns were taken at room temperature with the [120] incidence.
L and R represent the crystal chirality, $\CC= +$ and $-$, respectively, as determined by the comparison between the experimental and calculated patterns.
}
\end{center}
\end{figure*}
Figure 3 shows the typical over-focus Lorentz TEM images and CBED patterns obtained for single-crystalline domains of $\mnfegex$ (nominally $x$ = 1.0, 0.9 and 0.7).
The Lorentz TEM images shown in the left panels were taken well below the magnetic transition temperature ($\sim 100$ K) at zero field.
They show a stripy contrast at zero magnetic field due to the helical screw structure\cite{Uchida2006,Yu2010,Yu2011,Tonomura2012,Seki2012}.
Under the magnetic field ($\bm{B}$ = 100, 500, 200 mT, respectively) applied normal to the thin plate, these Lorentz TEM images change into spotty contrasts (the right panel) due to the formation of skyrmions\cite{Yu2010,Yu2011,Tonomura2012,Seki2012}.
In all single-crystalline domains, a fixed contrast of the skyrmions, either bright or dark, is observed, indicating the fixed helicity of skyrmion within the respective single-crystalline domain.
Image of skyrmions for $x$ = 1.0 shows two (upper and under) SkX domains with different contrasts; there is a grain boundary between different single-crystalline domains.
For $x \sim 0.9$, skyrmions are observed not in a crystalline state (SkX) but in an isolated form.
Nevertheless, they also show the fixed contrast within the single-crystalline domain.
Such a fixed in-plane configuration (either CW or CCW) suggests strong coupling between $\CC$ and $\MH$.

To determine $\CC$, we utilized the analysis of CBED, following the procedure developed by Tanaka {\it et al.}\cite{Tanaka1985}.
The reflection disks in the zeroth-order Laue zone (ZOLZ) reflections with the beam incident direction +[120] are shown in Fig. 3 (see Methods for the detailed indexing procedure.).
Asymmetric patterns in the 002 reflections and the $00\bar{2}$ ones due to $2_1$ crystal helix structure along the [001] direction were observed and used for determination of $\CC$.
The observed patterns of disks are compared with the simulated ones for each $\CC$, composition $x$ and thickness $t$ with use of a software MBFIT\cite{Tsuda1999}, as shown in Fig. 3.
In fact, the calculated CBED patterns assuming the fixed crystal chirality can well reproduce either of the experimentally observed patterns, enabling us to distinguish $\CC$ of the observed domain.
On these bases, we examined the correlation between $\CC$ and $\MH$ for other single-crystalline domains.
\begin{table}
\caption{
The crystalline chirality ($\CC$) and the magnetic helicity ($\MH$) as well as their correlation ($\CMC$) for domains of $\mnfegex$ with nominally in $x$ = 0.5, 0.7 and 1.0.
Each pair of $\CC$ and $\MH$ was determined in the same micro-crystal domain.
The $\MH$ for nominally $x$ = 0.9 was determined from isolated skyrmions in Lorentz TEM images.
The sign of $\CMC$ is reversed between $x = 0.7$ and $x = 0.9$, indicating the composition-induced reversal of $\CMC$.
}
\begin{tabular}{ccccc}
\hline
\ Composition $x$\ &\ Domain\ &\ \ \ $\CC$\ \ \ &\ \ \ $\MH$\ \ \ &\ \ $\CMC$\ \ \\
\hline
\hline
$\sim$ 0.5 & A & + & + & +\\
$\sim$ 0.5 & B & $-$ & $-$ & +\\
\hline
$\sim$ 0.7 & A & $-$ & $-$ & +\\
\hline
$\sim$ 0.9 & A & + & $-$ & $-$\\
$\sim$ 0.9 & B & $-$ & + & $-$\\
\hline
1.0 & A & + & $-$ & $-$\\
1.0 & B & $-$ & + & $-$\\
\hline
\end{tabular}
\end{table}
Table 1 shows the summary of $\CC$, $\MH$ and their correlation $\CMC$ as observed in each single-crystalline domain for $x = 1.0$ and $x \sim$ 0.5, 0.7, and 0.9.
Except for $x \approx 0.7$, we could observe domains of both $\CC$ to confirm that $\CMC$ is unique.
Remarkably, obtained $\CMC$ is opposite between the composition ranges of $x \le 0.5$ and $x \ge 0.7$.

\begin{figure}
\begin{center}
\ifTWOCOLUMN
\includegraphics[width=8cm]{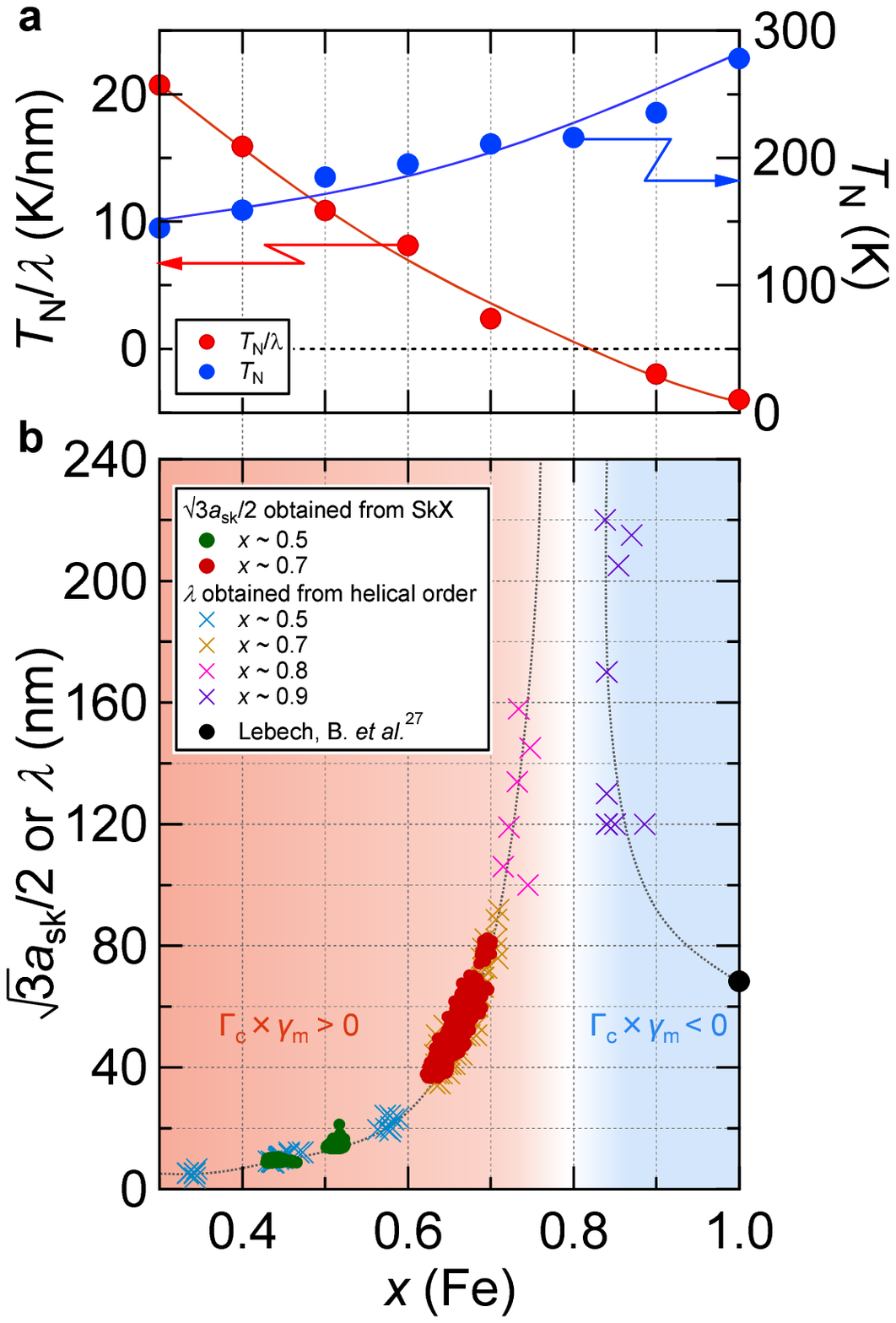}
\else
\includegraphics[width=10cm]{fig4.eps}
\fi
\caption{
{\bf Composition dependence of helical-magnetic properties of $\mnfegex$.}
{\bf a}, The $x$ (Fe) dependence of the helical magnetic transition temperature $T_\mathrm{N}$ and the quantity of $T_\mathrm{N}/\lambda$ (or $T_\mathrm{N} / \frac{\sqrt{3}\ask}{2}$) as the representatives for the symmetric and the antisymmetric (DM) exchange interactions, $J$ and $\alpha$, respectively (see Eq. (1)). Here, $\ask$ and $\lambda$ stand for the inter-skyrmion distance in SkX and the screw period, respectively.
Considering the sign change of $\CMC$, $T_{\rm{N}}/\lambda$ representing $\alpha$, is plotted as a negative value for $x > 0.8$.
{\bf b}, The $x$-dependence of magnetic periodicity,  $\frac{\sqrt{3}}{2}\ask$ or $\lambda$.
When the hexagonal SkX is represented as hybridization of three $q$-vectors of helix with the period $\lambda$, then the relation that $\frac{\sqrt{3}}{2}\ask = \lambda$ holds. 
This equality is ensured by the present observation.
The correlation between the crystal chirality and the magnetic helicity as represented by $\CMC$ is denoted with different background colors.
}
\end{center}
\end{figure}
Figure 4b shows the composition ($x$) dependence of the helical magnetic order period $\lambda$ or the corresponding SkX lattice constant $\ask$ multiplied with $\frac{\sqrt{3}}{2}$, as obtained from the Lorentz TEM analyses for the nominally $x$ = 0.5, 0.7, 0.8, 0.9, and 1.0 samples; the former four mixed-crystal specimens show spinodal decomposition with the composition spreading over the micro-domains and hence can provide the fine $x$-dependence data as already exemplified in Fig. 2c.
The determined sign of $\CMC$ is shown as background color in Fig. 4b.
Notably, a diverging behavior and a sign reversal of $\CMC$ are observed in the vicinity of $x = 0.8$. $\MH$ is related with the sign of DM interaction $\alpha$ in Eq. (1), 
while the sign of $\alpha$ should be determined by the signs of SOC and $\CC$. 
In a fixed-composition material, in which the sign of SOC is fixed, $\MH$ reversal occurs only via the inversion of $\CC$ (see Table 1).
In former studies\cite{Tanaka1985,Ishida1985} on MnSi, the effect of the DM interaction was calculated theoretically, and the result was confirmed to be consistent with the experimental results performed by CBED and polarized neutron diffraction.
The sign change of $\alpha$, or equivalently of the SOC, has been investigated on B20-type transition-metal silicides; Grigoriev {\it et al.}
determined $\CC$ and $\MH$ of the proper screw order with x-ray and polarized neutron diffractions, respectively, in $\mnfesix$ and $\fecosix$\cite{Grigoriev2009,Grigoriev2010}.
They showed that the $\CMC$ of $\mnfesix$ ($0 \le x \le 0.11$) and $\fecosix$ ($x$ = 0.10 and 0.25) are opposite to each other.
The sign change of $\alpha$ as observed from the skyrmion helicity in the present case may also reflect the change of the SOC sign in going from Mn-rich to Fe-rich B20 germanides as in the case of B20 silicides.

Here, we can further quantify the $\alpha$ change on the bases of the Lorentz TEM observations on the composition-spread samples.
Since the helical period $\lambda$ or almost equivalently the SkX lattice constant $\ask$ is proportional to $J/\alpha$ as derived from Eq. (1);
when $J \gg \alpha$ as in the present case, the magnetic transition temperature ($T_\mathrm{N}$) is in proportion to $J$.
Therefore, using the observed values of $T_\mathrm{N}$ and $\lambda$ or $\frac{\sqrt{3}}{2}\ask$, the relative change of $\alpha$ with the composition $x$ will be given by $T_\mathrm{ N} /\lambda$ (or $T_\mathrm{ N} / \frac{\sqrt{3}\ask}{2}$), while its sign  by $\CMC$.
We show in Fig. 4a the $x$ dependence of $T_\mathrm{ N}$ and $T_\mathrm{ N} /\lambda$ (or $T_\mathrm{ N} / \frac{\sqrt{3}\ask}{2}$) as representations of $J$ and $\alpha$, respectively.
The DM interaction strength $\alpha$ as measured by $T_\mathrm{ N} /\lambda$ (or $T_\mathrm{ N} / \frac{\sqrt{3}\ask}{2}$) shows a steep but almost linear change with $x$, while crossing zero, {\it i.e.} changing the sign, around $x = 0.8$.
Around this composition, a diverging behavior of $\lambda$ is observed as seen in Fig. 4b, where the SkX with the relatively large $\ask$  ($>$ 100 nm) can hardly be formed but there exist isolated skyrmions (see also the middle panels for $x\sim 0.9$ in Fig. 3).
Perhaps, the weakened (compensated) DM interaction is amenable to other competing interactions such as magnetic anisotropy or disordered (pinning) potential, suppressing the formation of a regular SkX lattice.

The strength and the sign of $\alpha$ or SOC in the present mixed-crystal system $\mnfegex$ appear to depend not only on the difference between the local SOC between Mn and Fe state but also on the filling of the amalgamated conduction band in $\mnfegex$.
The $\alpha$ value in the mixed crystal smoothly changes with $x$ as observed in the $T_\mathrm{ N}/\lambda$ (or $T_\mathrm{ N} / \frac{\sqrt{3}\ask}{2})$ vs.
$x$ curve (Fig. 4a).
This is perhaps because $\lambda$ or $\ask$ is large enough as compared with the chemical lattice constant, {\it i.e.}
the length scale of the atomic (Mn, Fe) spatial variation.

Such an amalgamation behavior in the magnetic interactions is rather contrastive with the tendency of the spinodal decomposition in the real-space chemical lattice.
The presently observed change of the skrymion size $\ask$ and its helicity with the chemical composition demonstrates the possibility of the critical tuning of the DM interaction strength.
In turn, the mixed-crystal engineering will enable us to control the skyrmion density and helicity, paving a way to skyrmion-based magnetoelectronics.

\section*{Methods}
$\mnfegex$ polycrystalline samples were prepared by arc-melting of stoichiometric mixtures of Mn, Fe, and Ge, followed by high-pressure synthesis (4.5 GPa, $800\ {}^\circ\mathrm{C}$, 1 hour).
The phase purity was checked by powder x-ray diffraction.
The calculated lattice constants from the peaks of B20 crystals approximately obey the Vegard's law.
Thin-plate samples for TEM were prepared by mechanical thinning and subsequent Ar-ion milling.
The Lorentz TEM study was carried out for the real-space magnetic structure imaging using a 200 kV TEM equipped with a CCD camera (JEM-2100F).
Control of sample temperature was achieved by a cooling sample-holder with flow of liquid nitrogen.
Magnetic field perpendicular to the specimen plate (or parallel to incoming electron beam) was applied by controlling the objective-lens current of TEM.

Composition maps were obtained by EDX using a 200 kV TEM equipped with a scanning transmission electron microscope (STEM) (JEM-2010F) and EDX detector (Bruker X-Flash 5030 with a $30\ \mathrm{mm}^2$ silicon drift detector).
Thickness maps were obtained by EELS with a 300 kV TEM (Hitachi HF-3000S).

For determination of the crystal chirality, CBED was measured with the same microscope as Lorentz TEM.
The CBED patterns containing the first-order Laue zone (FOLZ) were obtained by the CCD camera and compared with the patterns calculated by a software MBFIT\cite{Tsuda1999}.
The convergent-beam electrons were incident in the [120] direction, and 
the $[2\bar{1}0]$ ($[\bar{2}10]$) direction were determined 
as the direction in which the weaker (stronger) reflection in FOLZ
perpendicular to the $\langle 001 \rangle$ direction is observed experimentally.
Then, we identified the [001] direction and indexed the 002 and $00\bar{2}$ reflections.


\begin{acknowledgments}
The authors would like to thank N. Nagaosa, S. Seki, T. Kurumaji, and Y. Okamura for helpful discussions.
This study was supported by the Grant-in-Aid for Scientific Research (Grant No. 24224009) from the MEXT, and by Funding Program for World-Leading Innovative R\&D on Science and Technology (FIRST Program).
\end{acknowledgments}
\section*{Author contributions}
K. S. synthesized the polycrystalline samples, prepared the TEM samples, carried out the Lorentz TEM observations, and applied the CBED method.
X. Z. Y. measured the EELS thickness map.
T. H. measured the EDX composition map. D. M. analyzed the CBED patterns.
N. K. and S. I. contributed to the synthesis of polycrystalline samples.
K. K. and Y. M. contributed to the studies of EELS and EDX.
Y. T. conceived the project and wrote the manuscript with K. S.
\section*{Competing financial interests}
The authors declare no competing financial interests.
\end{document}